\RequirePackage{fix-cm}
\documentclass[twocolumn]{svjour3}          
\smartqed  
\usepackage{graphicx}
\usepackage{bbding}
\usepackage{upgreek}
\usepackage{url}
\usepackage{color,xcolor}
\usepackage{amsmath}
\usepackage{bm}
\usepackage[colorlinks,
            linkcolor=blue,
            anchorcolor=blue,
            citecolor=blue,
            urlcolor=blue,
            ]{hyperref}
\usepackage{mathptmx}      
\usepackage{titletoc}
\usepackage{amssymb}
\usepackage{flushend}
\usepackage{subfigure}
\usepackage{booktabs}
\usepackage{setspace}
\usepackage{algorithm}
\usepackage{algorithmicx}
\usepackage{algpseudocode}

\emergencystretch=\maxdimen
\titlecontents{section}[0pt]{\addvspace{2pt}\filright}              {\contentspush{\thecontentslabel\ }}              {}{\titlerule*[8pt]{.}\contentspage}

\hypersetup{
    bookmarks=false,         
    colorlinks=true,       
    linkcolor=blue,          
    linkbordercolor={1 1 1}
}

\journalname{Acta Mech. Sin.}
\begin{document}
\rmfamily
\title{Self-adaptive loss balanced Physics-informed neural networks for the incompressible Navier-Stokes equations
}

\titlerunning{Short form of title}        

\author{          Zixue Xiang$^1$ \and Wei Peng$^2$ \and Xiaohu Zheng$^1$ \and Xiaoyu Zhao$^2$ \and Wen Yao$^{2,*}$
       }


\institute{{\Envelope} Zixue Xiang \at
            \email{xiangzixuebit@163.com} \at \at
		    $^{1}$ College of Aerospace Science and Engineering, National University of Defense Technology, No. 109, Deya Road, Changsha 410073, China \at \at
            $^{2}$ National Innovation Institute of Defense Technology, Chinese Academy of Military Science, No. 55, Fengtai East Street, Beijing 100071, China
           }
\date{\copyright {\it Acta Mechanica Sinica}, The Chinese Society of Theoretical and Applied Mechanics (CSTAM) 2020
}


\maketitle


\begin{abstract}
There have been several efforts to Physics-informed neural networks (PINNs) in the solution of the incompressible Navier-Stokes fluid. The loss function in PINNs is a weighted sum of multiple terms, including the mismatch in the observed velocity and pressure data, the boundary and initial constraints, as well as the residuals of the Navier-Stokes equations. In this paper, we observe that the weighted combination of competitive multiple loss functions plays a significant role in training PINNs effectively. We establish Gaussian probabilistic models to define the loss terms, where the noise collection describes the weight parameter for each loss term. We propose a self-adaptive loss function method, which automatically assigns the weights of losses by updating the noise parameters in each epoch based on the maximum likelihood estimation. Subsequently, we employ the self-adaptive loss balanced Physics-informed neural networks (lbPINNs) to solve the incompressible Navier-Stokes equations,\hspace{-1pt} including\hspace{-1pt} two-dimensional\hspace{-1pt} steady Kovasznay flow, two-dimensional unsteady cylinder wake, and three-dimensional unsteady Beltrami flow. Our results suggest that the accuracy of PINNs for effectively simulating complex incompressible flows is improved by adaptively appropriate weights in the loss terms. The outstanding adaptability of lbPINNs is not irrelevant to the initialization choice of noise parameters, which illustrates the robustness. The proposed method can also be employed in other problems where PINNs apply besides fluid problems.
\end{abstract}
\vspace{1 cm}

\section{Introduction}
\hspace{1em} Incompressible fluids ranging from laminar to turbulent flows are widespread in many disciplines, such as environmental science, energy development, hydrology, and hydrogeology. It is indispensable to simulate the fluid precisely flows in engineering applications, e.g., conduction of heat, a vibration of the\hspace{-1pt} circular\hspace{-1pt} membrane,\hspace{-1pt} and\hspace{-1pt} propagation\hspace{-1pt} of electromagnetic waves. The Euler equations, advection-dispersion equations, and Navier–Stokes equations govern\hspace{-1pt} the\hspace{-1pt} fluid\hspace{-1pt} dynamics problems, which belong to partial differential equations (PDEs). Numerical simulations on fluid systems quietly rely on solving PDEs with the computational fluid dynamics (CFD) approach, including the finite elements (FE), finite volumes (FV), and finite differences (FD) methods \cite{2003Finite,2018Finite,2009Finite}. These discretization-based methods approximate the solution of PDEs through their values at a set of grid points distributed over the spatial and temporal domain. However, mesh generation for the flows with complicated geometry is often expensive and time-consuming. There are differences between actual mathematical properties of PDEs and approximate difference equations computed with the derivatives of state variables. In particular, the CFD simulations are challenging to solve the ill-posed or moving boundary value problems. To solve inverse problems by CFD method, this firstly requires tedious data assimilation and has no guarantee of convergence \cite{2012A}. Hence, the use of CFD models for real-life applications and real-time predictions is limited. It is significant to develop an effective Navier-Stokes solver that could overcome the limitations mentioned above.

One surrogate modeling approach to rapidly attain the solutions of Navier-Stokes equations, such as velocity, the pressure is to build a surrogate model, which learns the initial and boundary constraints from data \cite{Wang2016Data,2016Reconstruction,Sean2017Data,2019Flowfield,2018Deepinverse}. Due to the breakthrough approximation capabilities of neural networks \cite{MachineLearning,2019Subgrid}, there have been several remarkable results in solving forward and inverse problems for fluid simulation \cite{Data-drivenheattransfer,2000Neuralboundary,HiddenFluidMechanics}, instead of using the classical numerical schemes. However, the successful reconstruction of a flow field using neural networks is relevant to sufficient training data. It is challenging to compensate for the incompleteness and sparsity of measurements in many cases, and additional information is required. Recently, there is much effort to investigate physics-based models \cite{PIGP,HiddenPhysicsModels,PIGAN}. The emphasis is on obtaining the best surrogate model constrained by physical laws specified as PDEs besides data. In particular, physics-informed neural networks (PINNs) introduced in \cite{raissi2017physics,raissi2019physics} have already become novel PDEs solvers. A series of remarkable results across problems in various fields such as machinery and engineering, including biomedical problems \cite{CardiacActivationMapping}, finance \cite{OptionPricing}.

It is suitable for PINNs together with automatic differentiation \cite{2015AutomaticML} that does not require mesh generation to solve fluid mechanics problems \cite{PIDiagnostics,largesimulation,ADE}. Jin et al. proposed NSFnets by considering the velocity-pressure (VP) and the vorticity-velocity (VV) formulation of the incompressible Navier-Stokes equations \cite{2020NSFnets}. Mao et al. used PINNs to precisely approximate high-speed aerodynamic flows. Specifically, PINNs were employed to learn the value of the unknown parameter in the state equation for the oblique wave problem \cite{highspeedflows}. Sun et al. proposed physics-constrained DNNs for surrogate modeling of incompressible fluid flows without any data to avoid expensive simulation experiments \cite{0Surrogate}. The performance of the physics-constrained data-free DL surrogate model is studied on several flow cases with two idealized vascular geometries. Besides, the overriding concern is that current PINNs lack uncertainty quantification of the solution from the randomness of data or the model architecture restriction. Hence, Sun et al. developed a Physics Constrained Bayesian Neural Network to simulate 2D vascular flows from sparse, noisy velocity measurements. Variational inference \cite{UQ2020review} was used to estimate the posterior distributions of the reconstructed flow \cite{2020Physics}. Zhu et al. raised a Physics-Constrained Deep Learning methodology that simulates high-dimensional porous media flows without labeled data. The quantification and interpretation of the predictive uncertainty are also provided \cite{2019Physics}. As for inverse heat transfer applications, Cai et al. employed PINNs to simulate two-dimensional heat transfer in flow past a cylinder without thermal boundaries, which is almost impossible to be solved by classical methods \cite{HeatTransfer}.

However, the convergence accuracy of baseline PINNs permanently reduced to about $10^{-2}\pm 10^{-3}$, which is a weakness of PINNs \cite{raissi2017physics,raissi2019physics}. Therefore, it is necessary to raise the precision of PINNs from the essence of the method. Incorporating the governing equations, initial, and boundary constraints into the loss functions is the main feature of the physical model. The training process that minimizes the multiple loss functions could be regarded as a multi-objective task, challenging for global optimization methods, such as Adam, Stochastic Gradient Descent (SGD), and L-BFGS \cite{SGD,2014Adam,2015Stochastic}. Numerical experiments demonstrate that the performance of PINNs is closely linked with the appropriate combination of each loss term. There is no question that workforce to tune weights always time-consuming, laborious, and prone to errors and omissions. Besides, if gradient descent methods optimize the multiple objectives composed with fixed weights, it is quite possible to attain a locally optimal solution \cite{elhamod2020cophypgnn}. So far, there have been several efforts to integrating sustainable, balanced learning of loss function in PINNs, such as the Neural Tangent Kernel(NTK) principle \cite{NTK}, the training gradients \cite{wang2020understanding}, and the network weights \cite{Self-Adaptive}. Alex et al. observed that the optimal weighting of Multi-Task Learning is relevant to the magnitude of the task's noise \cite{kendall2018multitask}. Furthermore, the noise collection describes the weight parameter for each loss term by establishing Gaussian probabilistic models. From a different perspective, we provide a principled way of combining multiple loss functions with learning these competing loss terms in PINNs simultaneously.

In this work, we propose a self-adaptive loss balanced method for PINNs (lbPINNs), which automatically updates weights for each loss term in each iteration. We establish Gaussian probabilistic models to define complex loss functions based on maximum likelihood inference. Using a dynamic noise parameter collection determines the uncertainty of loss during training. The optimal weight of each loss constant depends on the magnitude of the noise parameter. We observe that the loss would be paid a severer penalty when the noise was declining. The primary goal is to improve the approximation capabilities of PINNs. The basic principle in lbPINNs is to tune the observation noise configuration of the model with gradient descent methods together with the training process. The effectiveness and merits of the proposed method are demonstrated by investigating several laminar flows, including two-dimensional steady Kovasznay flow, two-dimensional unsteady cylinder wake, and three-dimensional unsteady Beltrami flow. Compared with the baseline PINNs, the results show that the absolute error of self-adaptive loss-balanced PINNs can permanently be reduced to $10^{-4}\pm 10^{-5}$ in the identical experimental condition. To prove the adaptability of the method, we investigate the influence of initial noise parameters in the loss function on the accuracy of lbPINNs. All performance is better than the baseline PINNs. 

The paper is organized as follows. We first provide a detailed introduction to recent related work focused on balancing loss terms in PINNs in section 2. Section 3 introduces the framework of physics-informed neural networks for the incompressible Navier-Stokes equations and the principle of the self-adaptive loss balanced method for PINNs. In Section 4, numerical results of the developed approach on several incompressible Navier-Stokes flows are presented. Section 5 concludes the paper.

\section{Related Work}
\hspace{1em} The original PINNs algorithm has successful applications in Navier–Stokes, stochastic PDEs \cite{PIGAN,FokkerPlanck,FBSNN,2019Learning,2019Quantifying}, and fractional PDEs \cite{fPINNs}. However, it has been argued that the convergence and accuracy of PINNs still of tremendous challenge. There is no doubt that PINNs are obtained to minimize loss functions defined by the sampled data and physical laws that add to prior knowledge of PDEs. Zhao et al. observed that the combination of multiple loss functions plays a significant role in the convergence of PINNs \cite{wight2020solving}. In particular, more work has been devoted to balancing the interplay among loss terms in PINNs automatically through the training process.  

Recently, wang et al. provided a learning rate annealing algorithm that uses the back-propagated gradient statistics in the training procedure \cite{wang2020understanding}. It is widely acknowledged that the behavior of PINNs during model training via gradient descent is still a vague issue. The problems of vanishing gradient and exploding gradient would limit the application of this method. Hence the Neural Tangent Kernel(NTK) Perspective was presented to understanding the training process for PINNs \cite{NTK}. A practical technique based on the NTK perspective that appropriately assigns weights to each loss item was proposed. Since the distribution of eigenvalues of the NTK never changes, the performance improvement is quietly slight. Shin et al. proved the convergence theory for data-driven PINNs and derived the Lipschitz Regularized loss to solving linear second-order elliptic and parabolic type PDEs \cite{2020On}. A method that updates the adaptation weights in the loss function concerning the network parameters was suggested \cite{Self-Adaptive}. In general, the Self-Adaptive PINNs were forced to meet physical constraints as far as possible by minimizing the loss, where the trainable weights increase corresponding to higher loss. Hence the procedure could be remarked as a penalty PDE-constrained optimization problem.  

Although there have been many studies to verify the effects of loss on generalization performance, the competitive relationship between physics objectives loss items is not considered. Bu et al. pointed out that it is crucial to tune the competing physics-guided (PG) loss functions at various neural network learning stages \cite{elhamod2020cophypgnn}. Therefore, two approaches named annealing and cold starting that affect the initial or last epochs were proposed to tune the trade-off weights of loss terms with some different characteristics. However, it would be tedious to choose the appropriate kind of sigmoid function, which primarily influences accuracy. By careful consideration of competitiveness and adaptability, we get inspiration from multi-task learning and provide a self-adaptive loss balanced method for PINNs, which use uncertainty to weigh multiple losses automatically. We mainly investigate the possibility of using lbPINNs to approximate the incompressible Navier-Stokes equations that model laminar and turbulent flows.

\section{Methods}
\subsection{Physics-informed neural networks for the Navier-Stokes equations}
\hspace{1em} The physics-informed neural networks have been widely used as a data-driven method for solving general nonlinear partial differential equations, such as the Burgers equation, Poisson equation, and Schrodinger equation \cite{raissi2017physics,raissi2019physics}. We introduce physics-informed neural networks for the incompressible three-dimensional Navier-Stokes, reflecting the basic mechanics of viscous fluid flow. We model the fluid dynamics by the three-dimensional NS equations with the Newtonian assumption:
\begin{equation}
\begin{aligned}
\frac{\partial \mathbf{u}}{\partial t}+(\mathbf{u} \cdot \nabla) \mathbf{u} &=-\nabla p+ \frac{1}{\operatorname{Re}} \nabla^{2} \mathbf{u} \quad \text { in } \Omega, \\
\nabla \cdot \mathbf{u} &=0 \quad \text { in } \Omega, \\
\mathbf{u} &=\mathbf{u}_{\Gamma} \quad \text { on } \Gamma_{D}, \\
\frac{\partial \mathbf{u}}{\partial n} &=0 \quad \text { on } \Gamma_{N} ,\\
\mathbf{u}(\boldsymbol{x},0)&=\mathbf{h}(\boldsymbol{x}), \quad \text { in } \Omega, \\
\end{aligned}
\label{Navier–Stokes equation1}
\end{equation}
where $\boldsymbol{x}=[x,y,z] \in \Omega$ and $t \in[0, T]$ denote space and time coordinates. Respectively, the Reynolds number $Re = U_{ref}D_{ref}/\upsilon$ is defined by kinematic viscosity $\upsilon$, characteristic length $D_{ref}$, and reference velocity $U_{ref}$ of the fluid. The continuity equation of an incompressible fluid is considered, which describes the conservation of fluid mass. $\Gamma_{D}$ and $\Gamma_{N}$ are the Dirichlet and Neumann boundaries of the computational domain. The scalar $\boldsymbol{h}(\boldsymbol{x})$ is the initial constraints.  $u(x,y,z,t)$, $v(x,y,z,t)$, and $w(x,y,z,t)$ represent the $x, y$, and $z$ component of the velocity field. $p(x,y,z,t)$ represents the pressure. 

Following the original work of \cite{raissi2017physics}, the solution of Navier-Stokes equations \eqref{Navier–Stokes equation1} could be approximated by neural network $\hat{\boldsymbol{u}}(\boldsymbol{x},t;\theta)$. There are some advanced deep learning architectures, like Fourier Network\cite{2020Fourier}, DGM \cite{2018DGM}, and sinusoidal representation network (Siren) \cite{2020Implicit}. In its simplest form, we consider the feed-forward fully connected neural network of depth $M$, which takes the input $x,y,z,t$ and denote the output of the $m-th$ layer as $\hat{\boldsymbol{u}}^{[m]}$. The neural network is defined as:
\begin{equation}
\begin{aligned}
\text { input layer:} \quad \hat{\boldsymbol{u}}^{[m]}=&x, \\
\text { hidden layers:} \quad \hat{\boldsymbol{u}}^{[m]}=&\sigma\left(W^{[m]} \hat{u}^{[m-1]}+b^{[m]}\right), \\ \quad \text { for } m=2,3, \cdots, M-1, \\
\text { output layer:} \quad \hat{\boldsymbol{u}}^{[M]}=& \left(W^{[M]} \hat{u}^{[M-1]}+b^{[M]}\right),
\end{aligned}
\label{MLP}
\end{equation} 
where $\sigma$ is the activation function including sigmoid, relu, and tanh \cite{2018Activation,2017Searching}. $W^{[m]}$ and $b^{[m]}$ represent the weights and biases at $m-th$ layer. All weight matrices and \hspace{-1pt} bias\hspace{-1pt} vectors\hspace{-1pt} could\hspace{-1pt} be\hspace{-1pt} denoted by \vspace{1mm} 
a parameter collection $\theta=\left\{W^{[m]}, b^{[m]}\right\}_{1 \leq m \leq M}$. 
Define the residuals as:
\begin{equation}
\begin{aligned}
\boldsymbol{f}_{1}(\boldsymbol{x},t;\theta):&=\frac{\partial \mathbf{u}}{\partial t}+(\mathbf{u} \cdot \nabla) \mathbf{u} + \nabla p- \frac{1}{\operatorname{Re}} \nabla^{2} \mathbf{u},\\
\boldsymbol{f}_{2}(\boldsymbol{x},t;\theta):&=\nabla \cdot \mathbf{u}.
\end{aligned}
\label{residuals}
\end{equation}

\begin{figure}[htbp]
	\centering
	\subfigure{
		\includegraphics[scale=0.35]{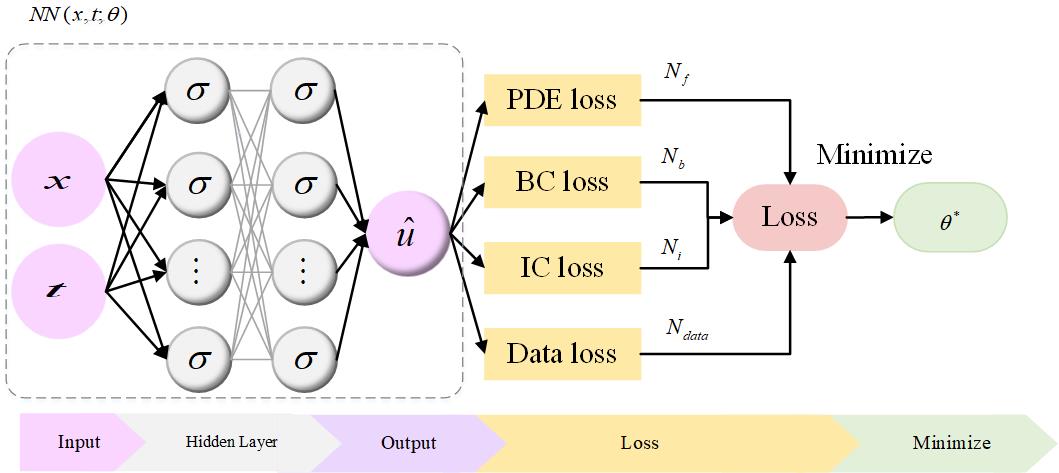}
	}
	\caption{A schematic diagram of the physics-informed neural networks (PINNs) for the incompressible Navier-Stokes equations.}
	\label{PINN}
\end{figure}

Automatic differentiation (AD) could take all partial derivatives, which has been integrated into PyTorch \cite{2015AutomaticML}.  Prior knowledge of physics was integrated into the loss function. The framework of physics-informed neural networks for the incompressible Navier-Stokes equations is constructed shown in figure \ref{PINN}. The best parameter collection $\theta ^{\ast }$ is identified by minimizing multiple loss functions as follows:
\begin{equation}
\begin{aligned}
L(\theta ; N)&=\omega_{f} L_{P D E}\left(\theta ; N_{f}\right)+\omega_{b}
L_{B C}\left(\theta ; N_{b}\right)\\ &+\omega_{i}
L_{I C}\left(\theta ; N_{i}\right)+\omega_{d} L_{d a t  a}\left(\theta ; N_{data}\right),
\end{aligned}
\label{PINNloss}
\end{equation}
where $\omega = \left\{\omega_{f},\omega_{b},\omega_{i},\omega_{d}\right\}$ contains weights of each loss term. The number of training points is concluded by parameter $N = \left\{N_{f},N_{b},N_{i},N_{data}\right\}$. The loss $L_{P D E}$ penalizes the Navier-Stokes residuals; The loss $L_{B C}$ aims to fit the Dirichlet boundary $\Gamma_{D}$ or Neumann boundary $\Gamma_{N}$. The loss $L_{I C}$ determines the error of the initial constraints. The loss $L_{d a t a}$ is computed corresponding to sample data:
\begin{equation}
\begin{aligned}
L_{P D E}\left(\theta ; N_{f}\right)&=\frac{1}{\left|N_{f}\right|} \sum_{j=1}^{N_{f}}\|\boldsymbol{f}_{1}(\boldsymbol{x}_{f}^{j}, t_{f}^{j}; \theta)\|^{2} + \|\boldsymbol{f}_{2}(\boldsymbol{x}_{f}^{j}, t_{f}^{j}; \theta)\|^{2}, \\
L_{B C}\left(\theta ;N_{b}\right)&=\frac{1}{\left|N_{b}\right|} \sum_{j=1}^{N_b}\|  \hat {\boldsymbol{u}}(\boldsymbol{x}_{b}^{j}, t_{b}^{j}; \theta) - \boldsymbol{g}_{b}^{j} \|^{2},\\
L_{I C}\left(\theta ;N_{i}\right)&=\frac{1}{\left|N_{i}\right|} \sum_{j=1}^{N_i}\|  \hat {\boldsymbol{u}}(\boldsymbol{x}_{i}^{j}, 0; \theta) - \boldsymbol{h}_{i}^{j} \|^{2},\\
L_{d a t a}\left(\theta ;N_{d a t a }\right)&=\frac{1}{\left|N_{d a t a}\right|} \sum_{j=1}^{N_{d a t a}}\|  \hat {\boldsymbol{u}}(\boldsymbol{x}_{data}^{j}, t_{data}^{j}; \theta) - \boldsymbol{u}_{data}^{j} \|^{2},
\end{aligned}
\label{eachloss}
\end{equation}
where $\left\{\boldsymbol{x}_{f}^{j}, t_{f}^{j}\right\}^{N_f}_{j=1}$ is a set of
collocation points that are uniformly sampled inside the domain $\Omega$. The collection $\left\{\boldsymbol{x}_{b}^{j}, t_{b}^{j}, \boldsymbol{g}_{b}^{j}=\boldsymbol{g}(\boldsymbol{x}_{b}^{j}, t_{b}^{j})\right\}^{N_b}_{j=1}$ donates the boundary constraint points. The collection $\left\{\boldsymbol{x}_{i}^{j}, \boldsymbol{h}_{i}^{j}\right\}^{N_i}_{j=1}$ donates the initial constraint points. The dataset $\left\{\boldsymbol{x}_{d a t a}^{j}, t_{d a t a}^{j},\boldsymbol{u}_{d a t a}^{j}\right\}^{N_{d a t a}}_{j=1}$ is sample points. The parameters $N_{f},N_{b},N_{i},N_{d a t a}$ are the total number of points. The Physics-informed neural networks (PINNs) for the incompressible Navier-Stokes equations algorithm is summarized as follows.

\begin{algorithm}[htb]
	\caption{Physics-informed neural networks (PINNs)}
	\label{alg:PINNs}
	\begin{algorithmic}
		\Require
		Training steps S, the learning rate $lr$, weight collection $\omega$ used to balance the interplay between the different loss terms. 
		\Ensure
		Find the best model with parameters $\theta ^{\ast }$ 
		\State \textbf{Step\hspace{0.5em}1}:Specify the training set \\
		Initial constraint points: $\left\{\boldsymbol{x}_{i}^{j}, \boldsymbol{h}_{i}^{j}\right\}^{N_i}_{j=1}$. \\
		Boundary constraint points: $\left\{\boldsymbol{x}_{b}^{j}, t_{b}^{j}, \boldsymbol{g}_{b}^{j}\right\}^{N_b}_{j=1}$.\\
		Training points: $\left\{\boldsymbol{x}_{d a t a}^{j}, t_{d a t a}^{j},\boldsymbol{u}_{d a t a}^{j}\right\}^{N_{d a t a}}_{j=1}$. \\
		Residual training points: $\left\{\boldsymbol{x}_{f}^{j}, t_{f}^{j}\right\}^{N_f}_{j=1}$.
		\State \textbf{Step\hspace{0.5em}2}: Construct the neural network $\hat{\boldsymbol{u}}(\boldsymbol{x},t;\theta)$ with random initialization of parameters $\theta$ \eqref{MLP}.
		\State \textbf{Step\hspace{0.5em}3}:Define the residuals \eqref{residuals} using automatic differentiation and other arithmetic operations.
		\State \textbf{Step\hspace{0.5em}4}:Use S steps of a gradient descent algorithm to update the parameters $\theta$ as:
		\For{$s = 1$ to $S$}
		\State (a) Specify the loss function $L(\theta_s ; N)$ \eqref{PINNloss} with the weight collection $\omega$ based on equation \eqref{eachloss}.
		\State (b) Update the parameters $\theta$ using Adam with the learning rate $lr$ for minimizing the loss function. \\
		\hspace{5em} $\theta_{s+1} \leftarrow$ Adam $(L(\theta_s ; N);lr)$
		\EndFor
	\end{algorithmic}
\end{algorithm}

\begin{table*}[!t]
	\begin{center}
		\caption{\label{tab1}Comparing the results with different weights constant when learning 2d Kovasznay flow with PINNs. The weight selection $[\omega_{f},\omega_{b}]=[0.25,0.75]$ achieves the best accuracy, which is bold in this table.}
		\begin{tabular}{cccccccccccc} \toprule
			Fixed weights  $\omega_{f},\omega_{b}$ &  $error_{u}$  &  $error_{v}$  &  $error_{p}$   \\ \hline
			$[1.0,1.0]$  & $4.06030e-03$  & $2.87294e-03$  & $4.43514e-03$ \\
			$[0.85,0.15]$  & $6.60475e-03$  & $4.69625e-02$  & $7.41679e-02$ \\
			$[0.15,0.85]$  & $3.83967e-03$  & $5.79027e-03$  & $3.84211e-02$ \\
			$[0.75,0.25]$  & $1.88834e-03$  & $1.97016e-03$  & $4.05601e-02$ \\
			\bm{$[0.25,0.75]$}  & \bm{$1.08440e-03$}  & \bm{$2.10914e-03$}  & \bm{$1.88593e-03$} \\
			$[0.65,0.35]$  & $8.46328e-03$  & $2.59047e-03$  & $1.97016e-02$ \\
			$[0.35,0.65]$  & $1.94430e-03$  & $3.70284e-03$  & $4.88638e-03$ \\
			\bottomrule
		\end{tabular}
	\end{center}
\end{table*}

\subsection{Self-adaptive loss balanced method}
\hspace{1em} The most common way to combine losses of each constraint is the weighted summation. To verify the impact of weight allocations, we simulated the 2d steady Kovasznay flow with different weight coefficients on the baseline PINNs. Here the weights are only considered $\omega_{f},\omega_{b}$ and add up to one. As the experimental results shown in table \ref{tab1}, we found that the performance and convergence of PINNs were susceptible to various loss weight selections. It is obvious that $[\omega_{f},\omega_{b}]=[0.25,0.75]$ achieve the best accuracy in this table. Besides, it could be expensive to manually tune these weight hyper-parameters, especially for complex fluid dynamics. Therefore, it is desirable to propose a more convenient and principled to learn these weights adaptively, achieving the balance between each competing physics objective.

\begin{figure}[htbp]
	\centering
	\subfigure{
		\includegraphics[scale=0.3]{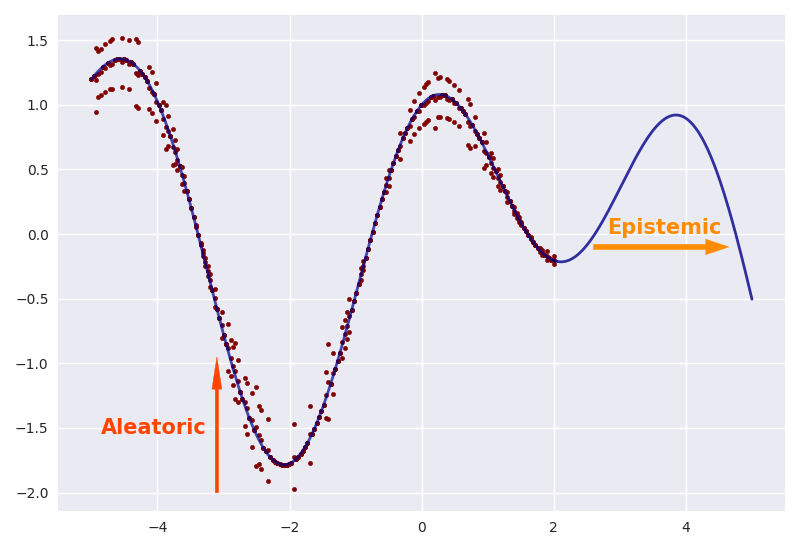}
	}
	\caption{View of the main differences between aleatoric and epistemic uncertainties. Aleatoric uncertainty captures the inherent property of data distribution and cannot be reduced even though more data were provided. Epistemic uncertainty indicates the uncertainty of the model and could be explained away when data are enough.}
	\label{UQ}
\end{figure}

Our idea is inspired by a paper that presents a principled strategy to weigh multiple loss functions in the scene geometry, and semantics multi-task deep learning problem \cite{kendall2018multitask}. The loss function is defined based on maximizing the Gaussian likelihood with the homoscedastic uncertainty. As is illustrated in figure \ref{UQ}, let us claim that the sources of uncertainty in Bayesian modeling are divided into two parts \cite{kendall2017uncertainties}. The first one, conceptually named epistemic uncertainty (also known as knowledge uncertainty), indicates the uncertainty of the model. Inadequate training data and knowledge always cause it. Thus it could be explained away when data are enough. Another type of uncertainty is aleatoric uncertainty (also known as data uncertainty), which captures the inherent property of data distribution and cannot be reduced even though more data were provided. Further, there are two subcategories of aleatoric uncertainty. The heteroscedastic uncertainty depends on the input data. It shows better performance when parts of the observation space have high noise levels. At the same time, homoscedastic uncertainty would stay constant for various inputs and varies between different tasks. 

As for regression tasks on solving Navier-Stokes equations, the likelihood is defined as a Gaussian with mean given by the surrogate model output $\hat {\boldsymbol{u}}(\boldsymbol{x},t;\theta)$. The parameter $\varepsilon$ is the observation noise in the output to determines the uncertainty:
\begin{equation}
\begin{array}{c}
p\left(y \mid \hat {\boldsymbol{u}}(\boldsymbol{x},t;\theta)\right)=N\left(\hat {\boldsymbol{u}}(\boldsymbol{x},t;\theta), \varepsilon^{2}\right).
\label{likelihood}
\end{array}
\end{equation}

\begin{figure}[htbp]
	\centering
	\subfigure{
		\includegraphics[scale=0.35]{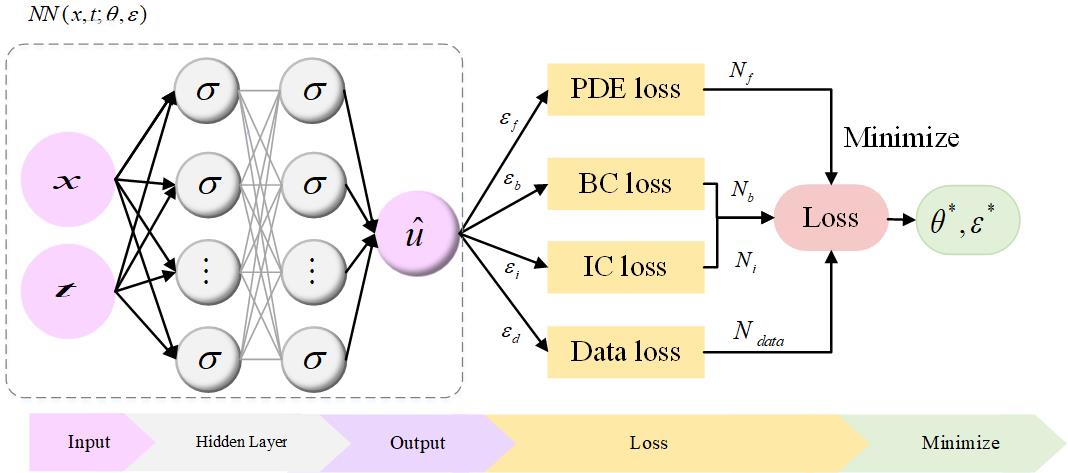}
	}
	\caption{Illustration of Self-adaptive loss balanced Physics-informed neural networks (lbPINNs).}
	\label{lbPINN}
\end{figure}

Consider the output follows Gaussian distribution. The noise scalar $\varepsilon$ is often fixed as part of the weight decay of neural networks. To capture aleatoric uncertainty with dependent data, we would tune the observation noise parameter based on maximum likelihood inference. Due to the minimization objective, the negative log-likelihood of the model is written as\cite{kendall2017uncertainties}:
\begin{equation}
\begin{aligned}
-\log p\left(y \mid \hat {\boldsymbol{u}}(\boldsymbol{x},t;\theta)\right) & \propto \frac{1}{2 \varepsilon^{2}}\left\|y-\hat {\boldsymbol{u}}(\boldsymbol{x},t;\theta)\right\|^{2}+\log \varepsilon \\
&=\frac{1}{2 \varepsilon^{2}} L_{1}(\theta)+\log \varepsilon,
\label{loglikelihood}
\end{aligned}
\end{equation}
where the loss $L_{1}(\theta)$ represents the output variable. Equation \eqref{loglikelihood} is composed of the residual regression and uncertainty regularization term to adapt the residual weights automatically. This idea is extended to the loss function of PINNs and read as follows:
\begin{equation}
\begin{aligned}
L(\varepsilon ; \theta ; N) & =  \frac{1}{2 \varepsilon_{f}^{2}} L_{P D E}\left(\theta ; N_{f}\right)+\frac{1}{2 \varepsilon_{b}^{2}} L_{B C}\left(\theta ; N_{b}\right)\\ & +\frac{1}{2 \varepsilon_{i}^{2}} L_{I C}\left(\theta ; N_{i}\right)+\frac{1}{2 \varepsilon_{d}^{2}} L_{d a t a}\left(\theta ; N_{d a t a}\right)\\ & +\log \varepsilon_{f} \varepsilon_{b} \varepsilon_{i} \varepsilon_{d},
\label{noise loss}
\end{aligned}
\end{equation} 
where the collection $\varepsilon = \left\{\varepsilon_{f}, \varepsilon_{b}, \varepsilon_{i}, \varepsilon_{d}\right\}$ describes noise parameters for each loss term. Similarly, the objective is to find the best model weights $\theta ^{\ast }$ and noise scalar $\varepsilon ^{\ast }$ by minimizing the loss $L(\varepsilon; \theta; N)$  with gradient-based optimizers, such as Stochastic Gradient Descent (SGD), Adam, and L-BFGS \cite{SGD,2014Adam,2015Stochastic}. On the one hand, for instance, $\varepsilon_{f}$ decreases, the weight of $ L_{P D E}$ increases, which means more punitive to the $ L_{P D E}$. The last term $\log \varepsilon_{f}$ regulates the $\varepsilon_{f}$. Thus the noise is discouraged from increasing too much. On the other hand, large uncertainty decreases the contribution of terms, whereas small scale would increase its contribution and penalizes the model. Moreover, the illustration of lbPINNs is shown in figure \ref{lbPINN}. We learn the relative weight of loss items adaptively when we optimize the noise scalar $\varepsilon$ and the model vector $\theta$, which allows us to learn the weights of each loss in a principled and persuasive way. The Self-adaptive loss balanced method for the PINNs algorithm is summarized.

\begin{algorithm}[htb]
	\caption{Self-adaptive loss balanced method for PINNs}
	\label{alg:lbPINNs}
	\begin{algorithmic}
		\Require
		Training steps S, the learning rate $lr$, initial values for the noise collection $\varepsilon = \left\{\varepsilon_{f}, \varepsilon_{b}, \varepsilon_{i}, \varepsilon_{d}\right\}$. 
		\Ensure
		Find the best model with parameters $\theta ^{\ast }$ and the excellent noise scalar $\varepsilon ^{\ast }$.
		\State \textbf{Step\hspace{0.5em}1}:Consider a physics-informed neural network $\hat{\boldsymbol{u}}(\boldsymbol{x},t;\theta)$ with initial parameters $\theta$.
		\State \textbf{Step\hspace{0.5em}2}:Construct a Gaussian probabilistic model \eqref{likelihood} with mean given by the output of PINNs and the noise collection $\varepsilon$.
		\State \textbf{Step\hspace{0.5em}3}:Then use S steps of a gradient descent algorithm to update the parameters $\varepsilon$ and $\theta$ as:
		\For{$s = 1$ to $S$}
		\State (a) Define the weighted loss function $L(\varepsilon_s ; \theta_s ; N)$ \eqref{noise loss} based on the maximum likelihood estimation \eqref{loglikelihood}.
		\State (b) Tune the noise parameters $\varepsilon$ via Adam to maximize the probability of meeting constraints.\\
		\hspace{5em} $\varepsilon_{s+1} \leftarrow$ Adam $ (L(\varepsilon_s ; \theta_s ; N);lr)$
		\State (c) Update the parameters $\theta$ via Adam.\\
		\hspace{5em} $\theta_{s+1} \leftarrow$ Adam $ (L(\varepsilon_s ; \theta_s ; N);lr)$
		\EndFor
	\end{algorithmic}
\end{algorithm}

\section{Results}
\hspace{1em} In this section, we test the original PINNs with different weight selections $\omega$ for the sake of contrasting their capability to represent manifold equations with lbPINNs. We aim to highlight the ability of our method to handle the classical Navier–Stokes equations, which are closely related to the physics of many scientific phenomena. The completed Navier–Stokes equation is often valid for engineering interests, such as vascular flow studies, Molecular diffusion analysis, airplane, and automobile design. We apply the proposed lbPINNs to simulate different incompressible Navier-Stokes flows. First, we consider two-dimensional steady Kovasznay flow with the analytic solution to investigate the effectiveness of lbPINNs based on boundary constraints. Then we employ the self-adaptive loss balanced method to unsteady cylinder wake in two dimensions. Finally, the three-dimensional unsteady Beltrami flow is also successfully simulated by lbPINNs, which considers initial and boundary constraints. CFD simulations provide all collocation points \cite{raissi2019physics}.

To illustrate the efficiency of the proposed method, the accuracy of the trained model is assessed through the relative L2 error of the exact value $\boldsymbol{u}(\boldsymbol{x}_{i}, t_{i})$ and the trained approximation $\hat {\boldsymbol{u}}\left(\boldsymbol{x}_{i}, t_{i}\right)$ inferred by the network at the data points $\left\{\boldsymbol{x}_{i}, t_{i}\right\}^{N}_{i=1}$. Various comparisons and detailed numerical experimental results on the Navier–Stokes equation are provided as follows.
\begin{equation}
\text { L2 error }=\frac{\sqrt{\sum_{i=1}^{N}\left|\hat {\boldsymbol{u}}\left(\boldsymbol{x}_{i}, t_{i}\right)-\boldsymbol{u}\left(\boldsymbol{x}_{i}, t_{i}\right)\right|^{2}}}{\sqrt{\sum_{i=1}^{N}\left|\boldsymbol{u}\left(\boldsymbol{x}_{i}, t_{i}\right)\right|^{2}}}.
\end{equation}

\begin{table*}[!t]
	\begin{center}
		\caption{\label{tab2}Comparing the results with multiple initial noise configurations when learning 2d Kovasznay flow.}
		\begin{tabular}{cccccccccccc} \toprule
			Initial settings  &  Excellent settings  &  $Loss_{PDE}$  &  $Loss_{BC}$  \\ \hline
			$[0.02,0.02]$  & $[2.476e-02, 9.472e-02]$  & $4.0734e-04,$  & $6.006e-03$ \\
			$[0.2,0.2]$  & $[1.866e-02, 1.996e-02]$  & $1.070e-04$  & $5.322e-03$ \\
			$[2,2]$  & $[1.945e-02, 8.095e-02]$   & $1.619e-04$  & $2.332e-03$  \\
			
			\bottomrule
		\end{tabular}
	\end{center}
\end{table*}

\subsection{Kovasznay flow}

\hspace{1em} In this example, we aim to emphasize the proposed methods to simulate the 2d steady Kovasznay flow that belongs to the incompressible Navier-Stokes flows. The analytic solution of velocity and pressure are given as follows \cite{2010An}:
\begin{equation}
\begin{aligned}
u(x, y)=&1-e^{\lambda x} \cos (2 \pi y), \\
v(x, y)=&\frac{\lambda}{2 \pi} e^{\lambda x} \sin (2 \pi y), \\
p(x, y)=&\frac{1}{2}\left(1-e^{2 \lambda x}\right), \\
where \quad \lambda=&\frac{1}{2 \nu}-\sqrt{\frac{1}{4 \nu^{2}}+4 \pi^{2}}, \quad \nu=\frac{1}{\operatorname{Re}}=\frac{1}{40}.
\end{aligned}
\end{equation}

The computational domain is $[-0.5, 1] \times [-0.5, 1.5]$. Let us consider the prediction scenario in which the 4-layer network is fully connected with 50 neurons per layer and a hyperbolic tangent activation function. There is no initial constraint for this steady flow. Specifically, the number of points selected for the trials shown is $N_b = 101, N_f = 2601$. The self-adaptive loss function is combined by PDE and boundary loss with $\varepsilon=[\varepsilon_{f},\varepsilon_{b}]$. We train this network for 1k epochs by minimizing the multiple loss equations using Adam optimizer with a learning rate of $0.001$. First, to explore the influence of $ [\varepsilon_{f}, \varepsilon_{b}]$ with different initial settings, we take several sets of initial noise configurations in which $ \varepsilon_{f} = \varepsilon_{b}$. The results, including the excellent settings, PDE, and boundary loss, are demonstrated in table \ref{tab2}. We find that the final configuration of $\varepsilon$ always achieves $10^{-2}$. The error of PDE residual and boundary is in the range $10^{-3}\pm 10^{-4}$. 

\begin{figure}[htbp]
	\centering
	\subfigure{
		\includegraphics[scale=0.45]{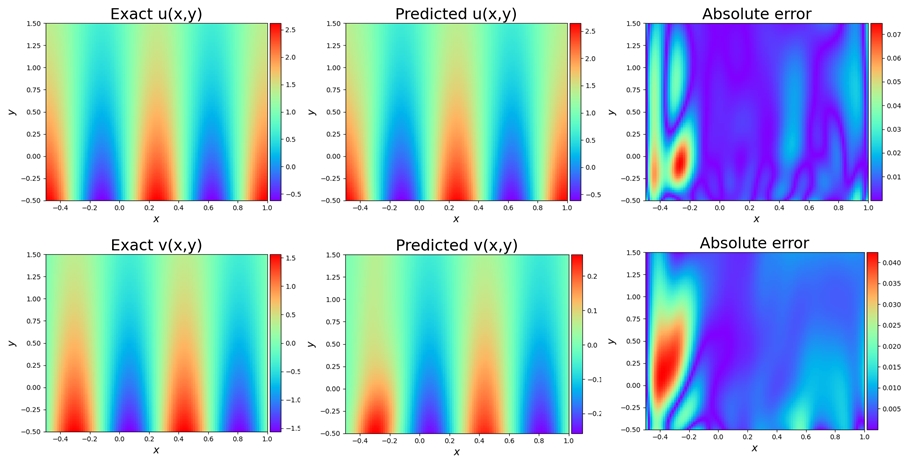}
	}
	\caption{The exact velocity $u(x,y)$, $v(x,y)$ (left), lbPINNs solution (middle), absolute error(right) for the two-dimensional Kovasznay flow across the space domain.}
	\label{2dK1}
\end{figure}

\begin{figure}[htbp]
	\centering
	\subfigure{
		\includegraphics[scale=0.42]{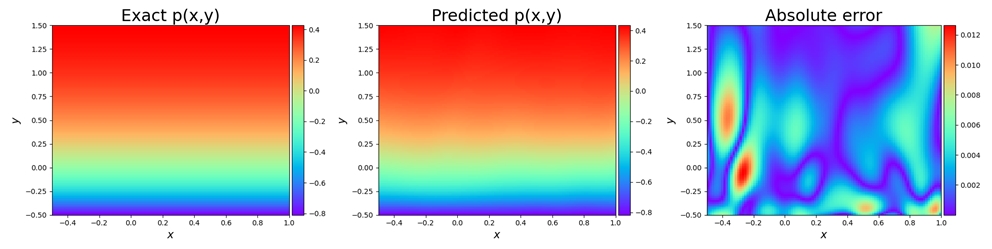}
	}
	\caption{Contrast exact solution(left) of pressure $p(x,y)$ for the two-dimensional Kovasznay flow with the result obtained by lbPINNs (middle) based on the numerical error (right).}
	\label{2dK2}
\end{figure}

\begin{figure}[htbp]
	\centering
	\subfigure{
		\includegraphics[scale=0.42]{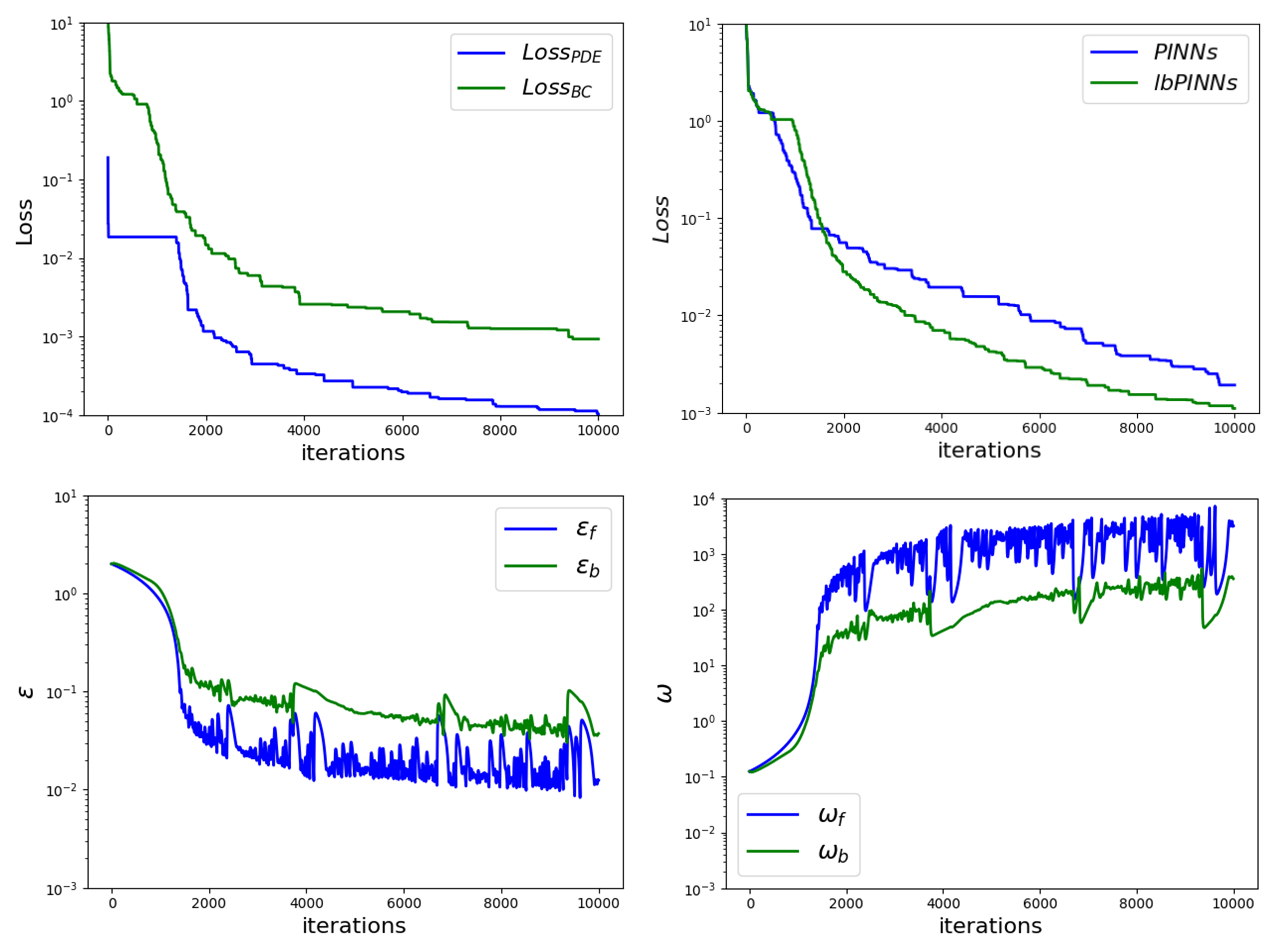}
	}
	\caption{$Loss_{PDE}, Loss_{BC}$, $\varepsilon_{f}, \varepsilon_{b}$ and $\omega_{f}, \omega_{b}$ diagrams for the two-dimensional Kovasznay flow are shown. Besides, the convergence of PINNs and lbPINNs are displayed.}
	\label{2dK3}
\end{figure}

Setting initial collection as $ [\varepsilon_{f}, \varepsilon_{b}] = [2, 2]$. The number of the test dataset is $101$. Finally, We achieved an L2 error of $6.411\times10^{-4}\pm4.320\times10^{-5}$ after 1k iterations, which is lower than the final errors of the baseline PINNs $4.435\times10^{-3}\pm 2.872\times10^{-3}$ over the same epochs. Figure \ref{2dK1} summarizes the predicted solution and the comparison with actual solutions of velocity $u(x,y)$, $v(x,y)$. We present the approximation of pressure $p(x,y)$ using lbPINNs and the exact solution in figure \ref{2dK2}. The relative error of the velocity and pressure for PINNs and lbPINNs are illustrated in table \ref{tab3}. We observe that the error of $u, v$, and $p$ achieve $10^{-4}\pm 10^{-5}$, demonstrating the effectiveness of dynamic weights. Additionally, the convergence of lbPINNs is more quickly than PINNs. The convergence of loss, noise constants, and weights during the training process are displayed in figure \ref{2dK3}. The most efficient noise configurations are in range $[5.102\times10^{-2}\pm1.945\times10^{-2}, 4.075\times10^{-2}\pm 2.127\times10^{-2}]$. Finally, all weights achieve $10^{3}$. The convergence of PDE and boundary loss depends on the dynamic noise configurations. The scalar $\varepsilon_{f}$ decreases rapidly and more punitive to the loss $ L_{P D E}$, which leads to faster convergence. The test loss of PINNs would attain $1.923\times10^{-3}$, while lbPINNs would converge to $1.106\times10^{-3}$.

\begin{table*}[!t]
	\begin{center}
		\caption{\label{tab3}Comparing the error for the velocity and pressure when learning 2d Kovasznay flow by PINNs with different weight selections $[\omega_{f},\omega_{b}]$ and lbPINNs with different initial noise configuration $ [\varepsilon_{f}, \varepsilon_{b}]$.}
		\begin{tabular}{cccccccccccc} \toprule
			Methods  &  parameter selections & $error_u$  &  $error_v$  &  $error_p$   \\ \hline
			\textbf{PINNs} &  \bm{$[1,1]$}  &\bm{$4.06030e-03$}  & \bm{$2.87294e-03$}  & \bm{$4.43514e-03$} \\
			PINNs &  $[10,10]$  &$5.18614e-03$  & $1.58971e-03$  & $2.50950e-03$ \\
			\textbf{lbPINNs}  &  \bm{$[2,2]$} & \bm{$6.41112e-04$}  & \bm{$2.81905e-04$}  & \bm{$ 4.32069e-05$} \\
			lbPINNs  &  $[0.2,0.2]$ & $2.15542e-04$  & $3.26974e-04$  & $ 3.27784e-05$ \\
			\bottomrule
		\end{tabular}
	\end{center}
\end{table*}

\begin{table*}[!t]
	\begin{center}
		\caption{\label{tab4}Apply multiple initial noise settings to simulate cylinder wake by lbPINNs.}
		\begin{tabular}{cccccccccccc} \toprule
			Initial settings  &  Excellent settings  &  $Loss_{PDE}$  & $Loss_{data}$   \\ \hline
			$[0.2,0.04]$  & $[1.447e-02, 8.240e-02]$  & $2.921e-04$ & $1.907e-04$ \\
			$[2,2]$  & $[9.964e-02, 4.421e-02]$   & $1.051e-03 $  & $7.882e-05$ \\
			$[2,0.04]$  & $[9.189e-02, 5.360e-02]$  & $6.987e-03$  & $6.828e-04$ \\
			
			\bottomrule
		\end{tabular}
	\end{center}
\end{table*}

\subsection{cylinder wake}
\hspace{1em} A two-dimensional incompressible flow and dynamic vortex shedding past a circular cylinder in a steady-state are numerically simulated using the spectral/hp element method to validate our predictions. Respectively,  the Reynolds number of the incompressible flow is $Re=100$. The kinematic viscosity of the fluid is $\upsilon = 0.01$. The cylinder diameter $D$ is $1$. The simulation domain size is $[-15,25]\times[-8,8]$, consisting of 412 triangular elements. Uniform velocity is imposed at the left boundary. Use the periodic boundary constraint on the top and bottom boundaries. A zero-pressure boundary is prescribed at the right boundary.

The deep neural network architecture is fully connected with layer sizes $[3, 20, 20, 20, 20, 20, 20, 20, 20, 2]$ and hyperbolic tangent nonlinearity. A physics-informed neural network model is trained 10k Adam iterations to approximate the latent solution $u(x,y,t), v(x,y,t)$, and $p(x,y,t)$ by formulating the composite loss. We sample $N_f = 4000$ collocation points, $N_{data} = 1000$ exact points. Numerical results of the Navier–Stokes equation in the way of the self-adaptive lbPINNs initialized with splendid initial noise settings are displayed in table \ref{tab4}. The final configuration of parameter $\varepsilon$ also always achieves $10^{-2}$. The error of PDE residual and data are in range $10^{-3}\pm 10^{-5}$. 

\begin{figure}[htbp]
	\centering
	\subfigure{
		\includegraphics[scale=0.42]{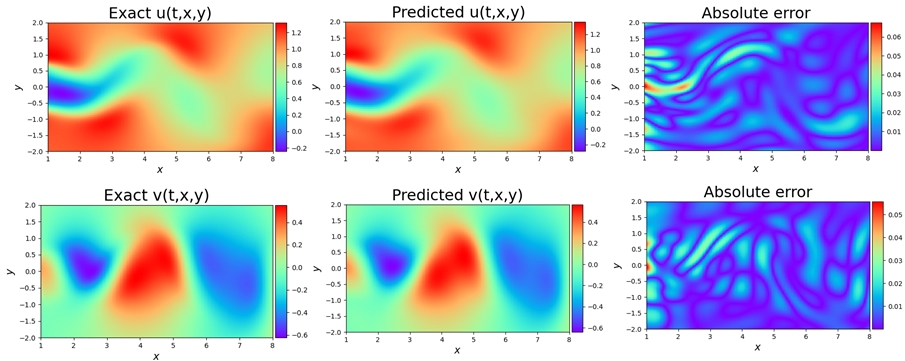}
	}
	\caption{Navier–Stokes equation:Contrast exact stream-wise velocity $u(x,y,t)$, (left) with the result obtained by lbPINNs (middle) based on the absolute error (right).}
	\label{2dNS1}
\end{figure}

\begin{figure}[htbp]
	\centering
	\subfigure{
		\includegraphics[scale=0.38]{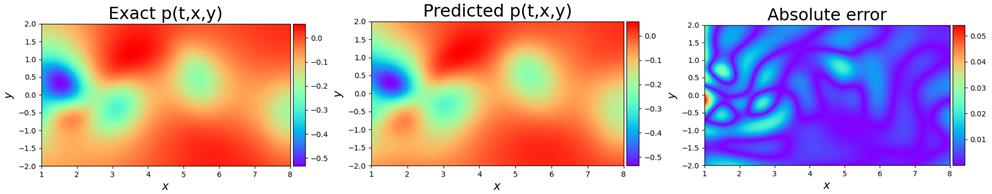}
	}
	\caption{Navier–Stokes equation:Contrast exact pressure $p(x,y,t)$ (left) with the result approximated by lbPINNs (middle) based on the absolute error (right).}
	\label{2dNS2}
\end{figure}

\begin{figure}[htbp]
	\centering
	\subfigure{
		\includegraphics[scale=0.42]{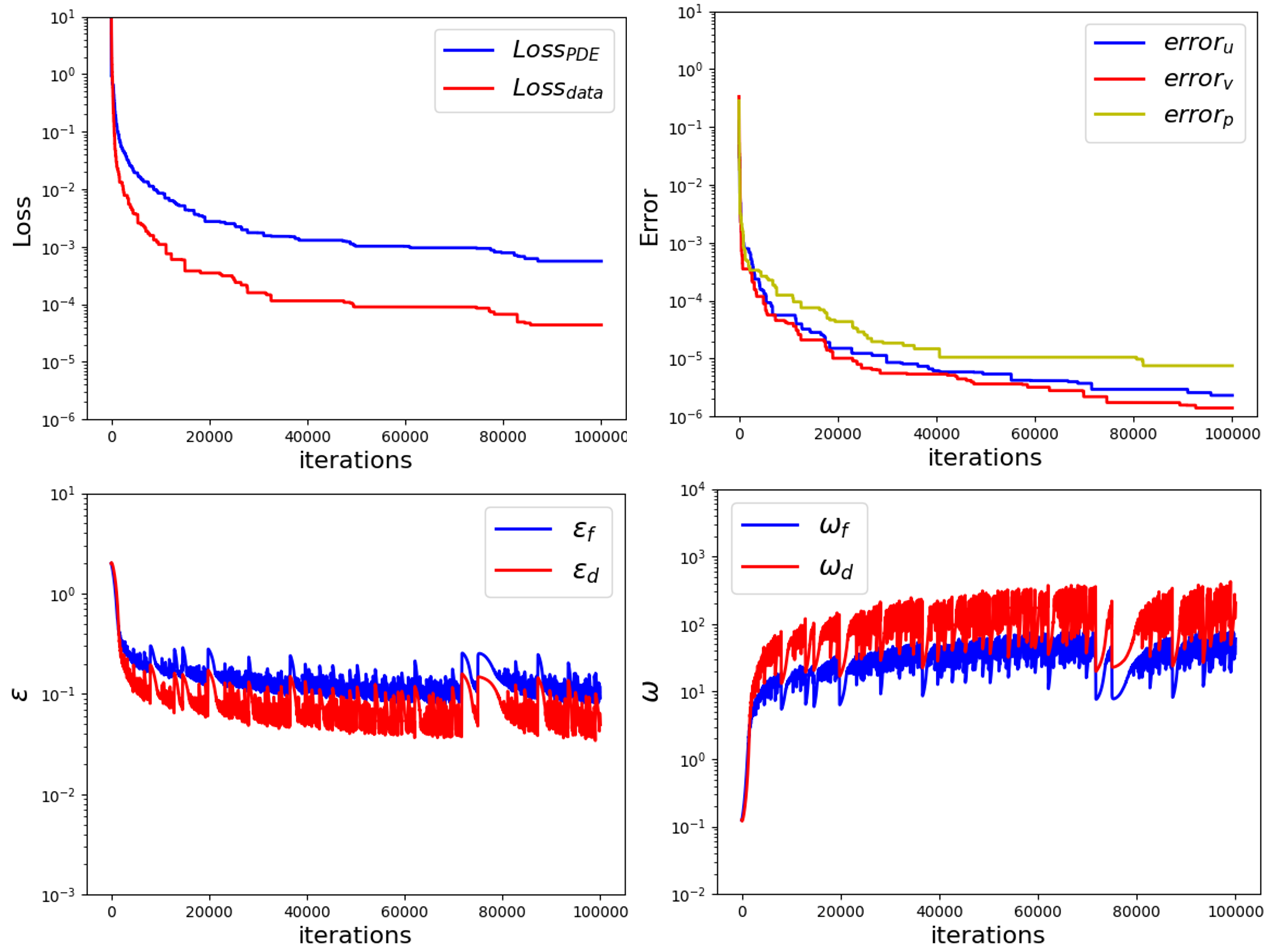}
	}
	\caption{$Loss_{PDE}, Loss_{BC}, Loss_{data}$, $\varepsilon_{f}, \varepsilon_{b}, \varepsilon_{d}$, and $\omega_{f}, \omega_{b}, \omega_{d}$ diagrams for the Navier–Stokes equations are shown. In addition, the convergence of $error_u, error_v$, and $error_p$ are displayed.}
	\label{2dNS3}
\end{figure}

We demonstrate the results of initial noise parameters $ [\varepsilon_{f}, \varepsilon_{d}] = [2,2]$ in detail. We present a comparison between the exact and the predicted solutions of stream-wise $u(x,y,t)$ and transverse velocity $v(x,y,t)$ at time instants $t = 4s$.  A more detailed assessment of the absolute error is displayed in the right panel of figure \ref{2dNS1}. In table \ref{tab5}, the resulting prediction error of velocity is measured at $4.818\times10^{-6}\pm 4.564\times10^{-6}$ in the relative L2-norm, while the original PINNs only converges to $5.321\times10^{-4}\pm 3.753\times10^{-4}$. That means extremely higher accuracy than the baseline PINNs. Based on the predicted versus instantaneous pressure field $p(x,y,t)$ shown in figure \ref{2dNS2}, the error between the predicted value and the true value is extremely low in the entire calculation domain. The convergence of $error_u, error_v$, and $error_p$ are shown in figure \ref{2dNS3}. Through observing the loss and noise parameters curves, we found that relative rates of loss items have evened out around 10k iterations, which indicates the customary efficiency of lbPINNs. It is necessary to demonstrate the adaptive procedure of two noise scalars, and the final result is $[9.964\times10^{-2}\pm 8.240\times10^{-2}, 7.388\times10^{-2}\pm 4.315\times10^{-2}]$. As expected, PINNs can accurately capture the complex nonlinear behavior of the Navier–Stokes equation with the self-adaptive loss function.  

\begin{table*}[!t]
	\begin{center}
		\caption{\label{tab5}Comparing the error for the velocity and pressure when simulating cylinder wake by PINNs with different weight selections $[\omega_{f},\omega_{d}]$ and lbPINNs with different initial noise configuration $ [\varepsilon_{f}, \varepsilon_{d}]$.}
		\begin{tabular}{cccccccccccc} \toprule
			Methods &  parameter selections &  $error_u$  &  $error_v$  &  $error_p$   \\ \hline
			\textbf{PINNs} &   \bm{$[1,1]$}& \bm{$5.32177e-04$}  & \bm{$3.75389e-04$}  & \bm{$5.26285e-03$} \\
			PINNs & $[10,10]$ & $2.30731e-04$  & $1.70630e-04$  & $3.19586e-03$ \\
			\textbf{lbPINNs} &  \bm{ $[2,2]$} & \bm{$4.56425e-06$}  & \bm{$4.81809e-06$}  & \bm{$ 1.06708e-04$} \\
			lbPINNs &  $[2,0.04]$ & $8.50794e-06$  & $6.71318e-06$  & $ 5.37089e-04$ \\
			
			\bottomrule
		\end{tabular}
	\end{center}
\end{table*}

\begin{table*}[!t]
	\begin{center}
		\caption{\label{tab6} Numerical results of Beltrami flow computed by lbPINNs initialized with Similar initial noise settings.}
		\begin{tabular}{cccccccccccc} \toprule
			Initial settings  &  Excellent settings & $Loss_{PDE}$  &  $Loss_{IC}$ &  $Loss_{BC}$ \\ \hline
			$[2,2,2]$  & $[2.260e-01, 1.171e-01, 3.140e-01]$ & $1.051e-02$  & $1.303e-03$ & $1.684e-02$\\
			$[0.2,2,0.2]$  & $[3.059e-02, 1.481962e-01, 3.549e-02]$ & $8.286e-02$  & $2.796e-02$ & $1.151e-02$ \\
			$[2,10,2]$  & $[2.718e-01, 1.705e+00, 6.625e-01]$ & $2.926e-02$  & $3.418e-02$ & $1.188e-02$ \\
			
			\bottomrule
		\end{tabular}
	\end{center}
\end{table*}

\subsection{Beltrami flow}
\hspace{1em} Applying the proposed algorithm to the three-dimensional Beltrami flow \cite{1994Exact}. The 3d unsteady Navier-Stokes flow has the following analytical solution:
\begin{equation}
\begin{aligned}
u(x, y, z, t)=&-a\left[e^{a x} \sin (a y+d z)+e^{a z} \cos (a x+d y)\right] e^{-d^{2} t}, \\
v(x, y, z, t)=&-a\left[e^{a y} \sin (a z+d x)+e^{a x} \cos (a y+d z)\right] e^{-d^{2} t}, \\
w(x, y, z, t)=&-a\left[e^{a z} \sin (a x+d y)+e^{a y} \cos (a z+d x)\right] e^{-d^{2} t}, \\
p(x, y, z, t)=&-\frac{1}{2} a^{2}\left[e^{2 a x}+e^{2 a y}+e^{2 a z}\right.\\
&+2 \sin (a x+d y) \cos (a z+d x) e^{a(y+z)}\\
&+2 \sin (a y+d z) \cos (a x+d y) e^{a(z+x)} \\
&\left.+2 \sin (a z+d x) \cos (a y+d z) e^{a(x+y)}\right] e^{-2 d^{2} t}
\end{aligned}
\end{equation}
where the parameters $a$ and $d$ are 1. The spatial domain is $[-1,1] \times [-1,1] \times [-1,1]$. The time interval is $[0,1]$. For this problem, we study the accuracy of the lbPINN method for the Reynolds number $Re=1$. We fix the network architecture with $layers= 10$ and $layer sizes = [4, 100, 100, 100, 100, 100, 100, 100, 100, 100, 100, 4]$. The total numbers of training points are maintained as $N_{b} = 100,  N_{i} = 100$, and $N_{f} = 2601$. Besides, the activation function is tanh and then train the network with a learning rate of 0.001 for 5000 epochs by optimizing equation \eqref{noise loss}. We also present the errors of simulation results of lbPINNs with different initial settings ${\varepsilon_{f}, \varepsilon_{i}, \varepsilon_{b}}$ in table \ref{tab6}. The final configuration of collection $\varepsilon$ also always achieves $10^{-1}$, which results in the error of PDE residual, boundary, and initial are in the range $10^{-2}\pm 10^{-3}$.

\begin{figure}[htbp]
	\centering
	\subfigure{
		\includegraphics[scale=0.40]{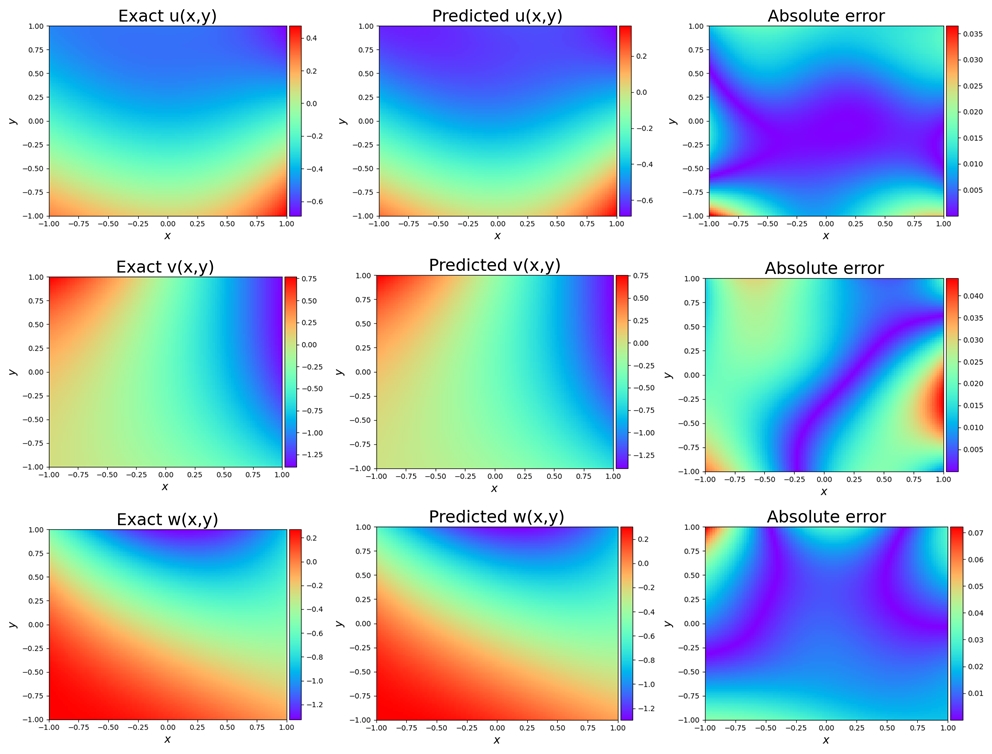}
	}
	\caption{According to temporal snapshots $t=1$ on the plane $z=0$ to compare the exact velocity $u(x, y)$ (Top), $v(x, y)$ (Middle), $w(x,y)$ (Bottom) and predicted solutions of three-dimensional Beltrami flow solved using the self-adaptive loss balance approach.}
	\label{B1}
\end{figure}

\begin{figure}[htbp]
	\centering
	\subfigure{
		\includegraphics[scale=0.40]{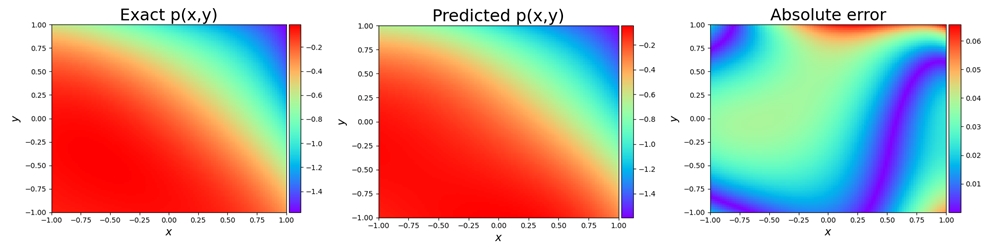}
	}
	\caption{Beltrami flow: analytical solutions and numerical approximations of pressure $p(x, y)$ at $t=1$ on the plane $z=0$ using the developed PINNs.}
	\label{B2}
\end{figure}

\begin{figure}[htbp]
	\centering
	\subfigure{
		\includegraphics[scale=0.42]{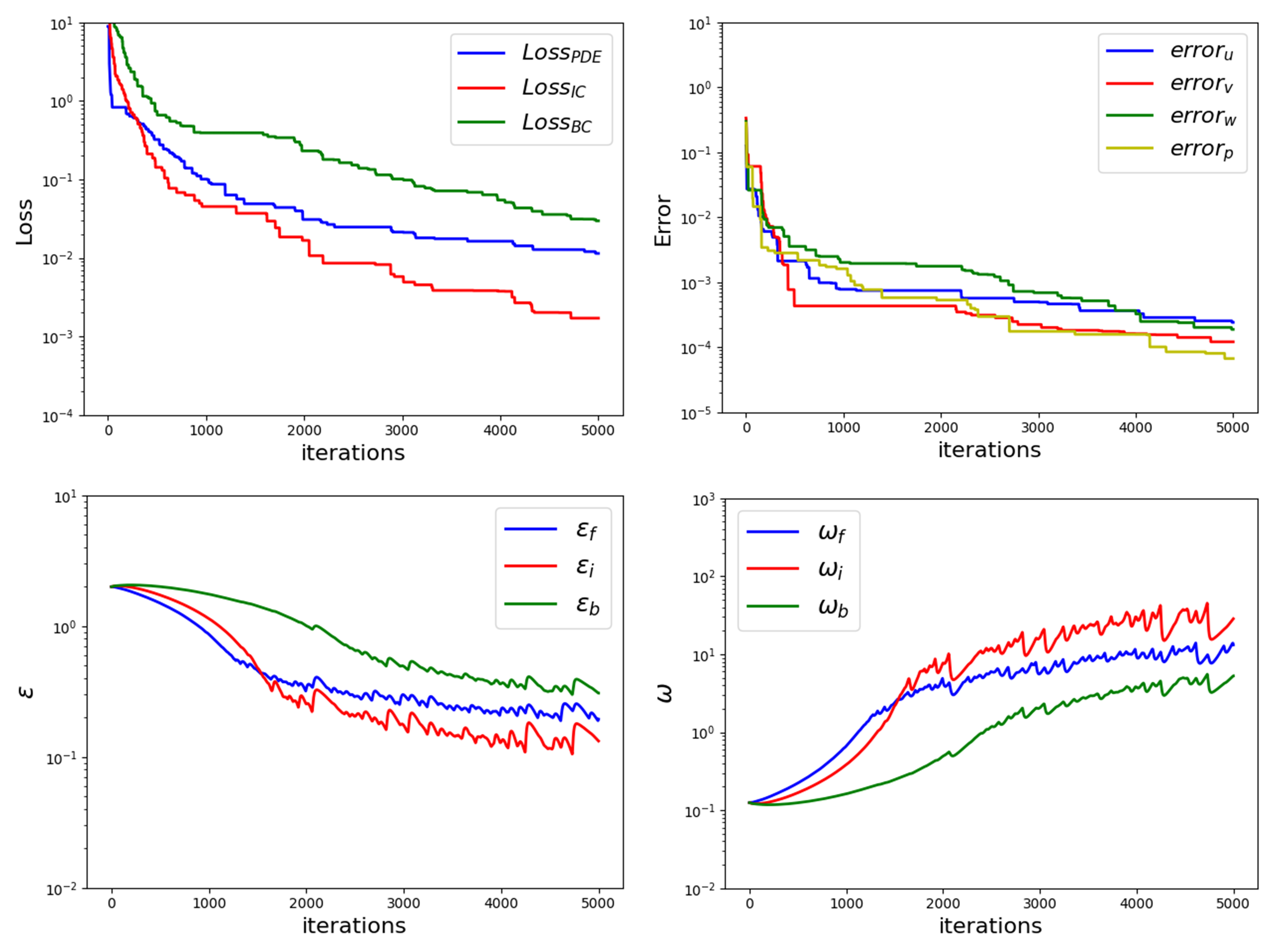}
	}
	\caption{$Loss_{PDE}, Loss_{BC}$, $\varepsilon_{f}, \varepsilon_{b}$, and $\omega_{f}, \omega_{b}$ diagrams for the two-dimensional Kovasznay flow are shown. The curves of $error_u, error_v, error_w$, and $error_p$ are also displayed.}
	\label{B3}
\end{figure}

We consider $ [\varepsilon_{f}, \varepsilon_{i}, \varepsilon_{b}] = [2,2,2]$ to its initial settings. In figure \ref{B1}, the relative absolute errors between the predicted and exact solution of velocity $u(x, y)$, $v(x, y)$, and $w(x,y)$ at a representative time instant on the plane $z=0$ are presented. In addition, we employ lbPINNs to predict the pressure $p(x,y)$ in figure \ref{B2}. In order to compare clearly, the average error of the velocity and pressure for PINNs and lbPINNs are illustrated in table \ref{tab7}. It is obvious that the predictive error computed with L2 error over 5000 epochs of velocity and pressure could attain $10^{-4}\pm 10^{-5}$, which denominates the high precision of self-adaptive loss balanced method. We further display the curves of $error_u, error_v, error_w$, and $error_p$ in figure \ref{B3}. The loss, noise constants, and weights number of iterations are also shown. The range of excellent noise parameters is $[2.614\times10^{-1}\pm 2.260\times10^{-1}, 1.965\times10^{-1}\pm1.171\times10^{-1}, 3.140\times10^{-1}\pm3.029\times10^{-1}]$. The parameter $\varepsilon_{b}$ came down slower than other parameters at the same iterations, which leads to lower $\omega_{b}$  and higher $Loss_{BC}$. 

\begin{table*}[!t]
	\begin{center}
		\caption{\label{tab7}Comparing the average absolute error of the velocity and pressure using baseline PINNs with different weight selections $[\omega_{f},\omega_{i},\omega_{b}]$ and lbPINNs with different initial noise configuration $ [\varepsilon_{f}, \varepsilon_{i}, \varepsilon_{b}]$and lbPINNs.}
		\begin{tabular}{cccccccccccc} \toprule
			Methods&  parameter selections   &  $error_u$  &  $error_v$  &  $error_w$  & $error_p$   \\ \hline
			\textbf{PINNs}&  	\bm{$[1,1,1]$}  & \bm{$1.18639e-02$}  & \bm{$2.95078e-03$}  & \bm{$ 1.39362e-03$}  &$1.04428e-02$ \\
			PINNs&  	$[20,10,2]$  & $7.86194e-03$  & $1.64679e-03$  & $ 7.65673e-03$  &$9.84695e-02$ \\
			\textbf{lbPINNs}&  	\bm{$[2,2,2]$}   & \bm{$3.21078e-04$}  & \bm{$9.90049e-05$}  & \bm{$2.34041e-04$}  & \bm{$ 1.87083e-04$} \\
			lbPINNs&  	$[2,10,2]$   & $1.82803e-04$  & $8.64701e-05$  & $5.61524e-04$  & $ 1.28763e-04$ \\
			\bottomrule
		\end{tabular}
	\end{center}
\end{table*}
\section{Conclusions}
\hspace{1em} In this paper, we observe that there exists a competitive relationship between complex physics loss items in PINNs. The performance and convergence of PINNs were susceptible to various loss weight selections. Thus it is beneficial to adaptively assign appropriate weights to combine multiple loss functions in PINNs. We propose a self-adaptive loss balanced method based on maximizing the Gaussian likelihood with the scalable uncertainty parameters to learn these competing loss terms in PINNs simultaneously. The objective of this work is to reconstruct the incompressible Navier-Stokes flows with higher accuracy. The effectiveness and merits of lbPINNs are demonstrated by investigating several laminar flows, including two-dimensional steady Kovasznay flow, two-dimensional unsteady cylinder wake, and three-dimensional unsteady Beltrami flow. Various experimental results could support our claim that the decay of the loss function is slightly faster, and the relative error is lower than that of original PINNs. Moreover, we initialize different sets of uncertainty parameters to indicate the outstanding adaptability of lbPINNs.

The self-adaptive loss balanced Physics-informed neural networks make progress in simulating unsteady Navier-Stokes flow field. However, the noise parameters that play a leading role in weighting loss items are determined by the optimization algorithms for training deep neural networks. To the best of our knowledge, there is no guarantee that gradient-based optimization algorithms would find the exact solution. It is worth attaching importance to theoretical analysis on the phenomenon and improving the robustness and scalability.

\section*{Acknowledgments}
This work was supported by the Postgraduate Scientific Research Innovation Project of Hunan Province (No.CX20200006) and the National Natural Science Foundation of China (Nos.11725211 and 52005505).

\end{document}